\documentstyle[prl,aps,floats,epsf]{revtex}

\newcommand {\Heb}        {\mbox{$\overline{\mbox{\rm He}}$}}

\newcommand {\prt}        {\mbox{$p$}}  
  
\newcommand {\pb}         {\mbox{$\overline{p}$}}

\begin{document}
\draft

%%%%%%%%%%%%%%%%%%%%%%%%%%%%%%%%%%%%%%%%%%%
\twocolumn[                               %
\hsize\textwidth\columnwidth\hsize\csname %
@twocolumnfalse\endcsname                 %
%%%%%%%%%%%%%%%%%%%%%%%%%%%%%%%%%%%%%%%%%%%

\title{A New Limit on the Flux of Cosmic Antihelium\footnotemark}
\author{
 T.~Saeki,$^{1}$  
 K.~Anraku,$^{1}$ 
 S.~Orito,$^{1}$  
 J.~Ormes,$^{5}$ 
 M.~Imori,$^{1}$
 B.~Kimbell,$^{3}$ Y.~Makida,$^{2}$  
 H.~Matsumoto,$^{4}$  H.~Matsunaga,$^{1}$  
 J.~Mitchell,$^{5}$  
 M.~Motoki,$^{4}$ 
 J.~Nishimura,$^{6}$  M.~Nozaki,$^{4}$ 
 M.~Otoba,$^{1}$ 
 T.~Sanuki,$^{1}$ 
 R.~Streitmatter,$^{5}$  J.~Suzuki,$^{2}$ 
 K.~Tanaka,$^{2}$ 
 I.~Ueda,$^{1}$ 
 N.~Yajima,$^{6}$  T.~Yamagami,$^{6}$ 
 A.~Yamamoto,$^{2}$ 
 T.~Yoshida$,^{2}$ 
 and K.~Yoshimura$^{1}$}

\address{
$^{1}$University of Tokyo, Bunkyo-ku, Tokyo 113, Japan
}
\address{
$^{2}$High Energy Accelerator Research Organization (KEK), 
Tsukuba, Ibaraki 305, Japan
}
\address{
$^{3}$New Mexico State University, Las Cruces, NM 88003, U.S.A
}
\address{
$^{4}$Kobe University, Kobe, Hyogo 657, Japan
}
\address{
$^{5}$National Aeronautics and Space Administration, Goddard Space Flight 
Center (NASA/GSFC), Greenbelt, MD 20771, U.S.A.
}
\address{
$^{6}$The Institute of Space and Astronautical Science (ISAS), Sagamihara, 
Kanagawa 229, Japan
}

\date{\today} 
\maketitle 

\begin{abstract} 
A very sensitive search for cosmic-ray antihelium was performed using data
obtained from three scientific flights of BESS 
magnetic rigidity spectrometer.
We have not observed any antihelium; this places a model-independent 
upper limit (95 \% C.L.) on the antihelium flux of
$6\times10^{-4}$ m$^{-2}$sr$^{-1}$s$^{-1}$
at the top of the atmosphere in the rigidity region 1 to 16 GV,
after correcting for the estimated interaction loss of antihelium in
the air and in the instrument.
The corresponding upper limit on the \Heb/He flux ratio
is 3.1 $\times 10^{-6}$, 30 times more stringent than
the limits obtained in similar rigidity regions
with magnetic spectrometers previous to BESS.
\end{abstract} 
%%%%%%%%%%%%%%%%%%%%%%%%%%%%%%%%%%%%%%%%%%%
]                                         %
%%%%%%%%%%%%%%%%%%%%%%%%%%%%%%%%%%%%%%%%%%%

\narrowtext
\footnotetext{\footnotemark 
This work is dedicated to the memory of Dr. R. Golden.}
Cosmic-ray observations provide most direct evidence for our Galaxy being 
composed mostly by baryons. This baryon-antibaryon asymmetry can be
global in the Universe, being created in the very early Universe due
to the violations of CP and of baryon-number.
However, depending on the nature of CP violation,
baryon-symmetric models are conceivable~\cite{BSYM} in which 
the Universe is separated 
into an equal number of matter- and antimatter-domains.
Whereas $\gamma$-ray observations place strong limitations on the antimatter 
in our Galaxy and in the local cluster of galaxies, 
the domain structure could still exist beyond this scale.
Although antihelium might be in principle produced in cosmic-ray interactions, 
their contribution to the \Heb/He flux ratio 
is expected to be much smaller than $10^{-12}$~\cite{EV72}.
Detecting antihelium at a level higher than this could therefore provide 
the evidence of antimatter domains or of other exotic phenomena
such as superconducting strings in our Galaxy~\cite{WT85}.
For further discussion of astrophysical considerations regarding the search 
for antihelium see the references~\cite{OR97,SGNF}.
We report in this letter (for more detail see~\cite{SA96}) 
a sensitive search for antihelium 
using the data from the '93, '94, and '95 flights of BESS detector,
and provide model-independent upper limits on
the absolute antihelium flux as well as on the \Heb/He flux ratio.
A limit on the flux ratio from
an early analysis of '95 data is published elsewhere~\cite{OR97}.

\begin{figure}[b]
\centerline{\mbox{\epsfxsize=17cm 
                  \epsffile{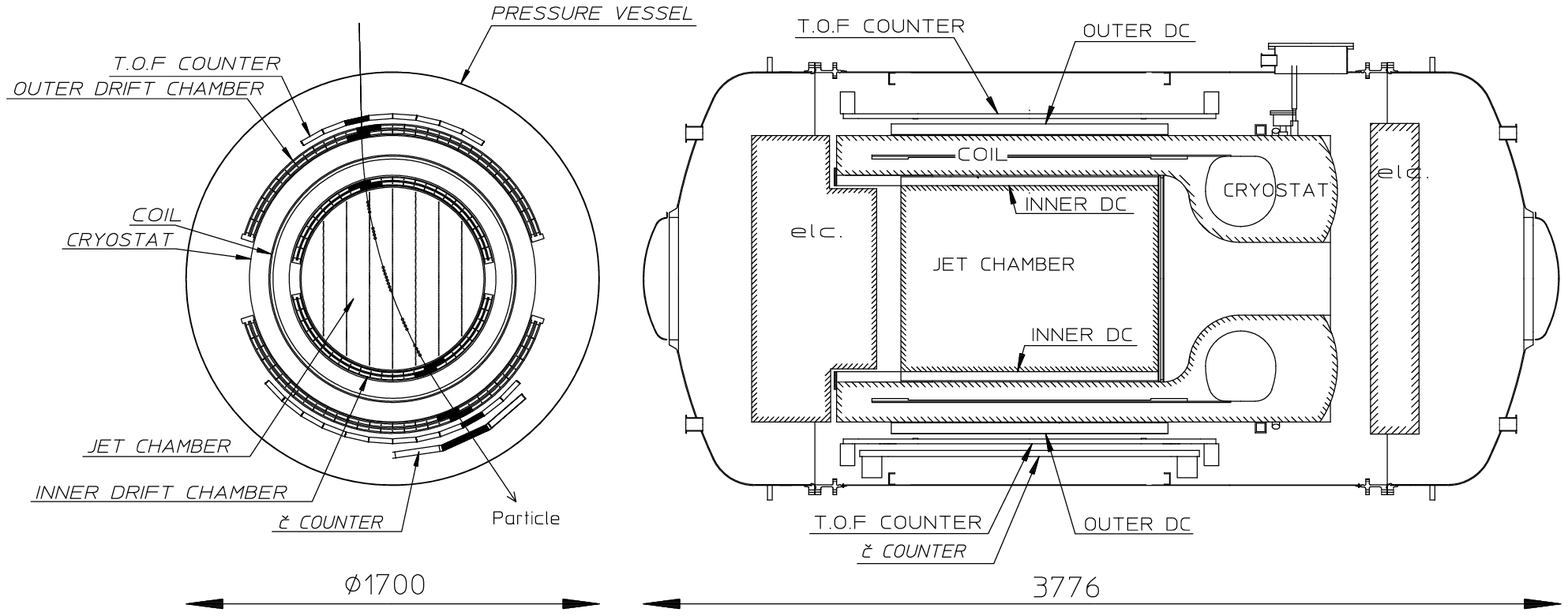}}}
\caption{Cross-sectional front- and side-views of the  BESS '95 instrument.}
\label{fig:bess}
\end{figure}

Fig.\ \ref{fig:bess} shows front- and side-views of the BESS '95 instrument.  
The cylindrical configuration provides a wide tracking region and an 
acceptance of up to 0.32 m$^2$sr depending on the off-line fiducial cuts.
From inside to outside, it includes a jet-type drift (JET) chamber,  
inner drift chambers (IDCs), a superconducting solenoid,  
outer drift chambers (ODCs), and a time of flight (TOF) hodoscope.
The solenoid produces a magnetic field of 1 T 
with an uniformity of $\pm 15\%$ inside the bore.

The JET chamber~\cite{drum80}, as the key tracking detector, 
measures up to 24 points per track three-dimensionally, each with 
the resolution~\cite{resol} of 200 $\mu$m in $r\phi$ plane and
of 2 cm in $z$-position
(a cylindrical coordinate ($r\phi z$) is determined 
by defining the magnetic field direction as $z$-axis).
Each of the IDCs and ODCs consists of two 12-mm-thick drift layers 
which are divided into 50-mm-wide cells, and measures 
$r\phi$ position each with 200 $\mu$m resolution.  
These redundant and continuous position measurements with ODCs, IDCs, and JET 
chamber, all equipped with multi-hit capacity, make it possible to 
recognize multi-track events and tracks having interactions and scatterings,
thus minimizing the background originating from the interactions.
The $r\phi$-tracking in the central region is performed by fitting hit points 
in the JET chamber as well as in the IDCs.
This results in a maximum detectable rigidity ($R$) of 200 GV and
a typical rigidity resolution of 0.5 \% at 1 GV.
Using vernier pads with a cycle of 10 cm for IDCs and 12 cm for ODCs, 
both chambers can measure the $z$ position, modulo the cycle, with 300 
$\mu$m resolution. 
Combining the IDC and JET information, 
$z$-position of the track can be determined with a precision of 300 $\mu$m.

The TOF hodoscope, placed at a radius of 65 cm, 
consists of eight upper and twelve lower plastic 
scintillators (Bicron 404), each of which has a dimension of 
95 cm $\times$ 10 cm $\times$ 2 cm.  
The light signals from the scintillator are guided 
through acrylic light-guides and reach
the photomultiplier-tubes (PMTs) attached on both ends. 
The timing and amplitude of the PMT signals are measured to determine the 
time of flight and d$E/$d$x$ of incident particles with resolution of 110 ps 
and 10 \%, respectively.
For '93 and '94 instruments, 
the TOF hodoscope consisted of four upper and six lower plastic 
scintillators, each of which had a dimension of 
110 cm $\times$ 20 cm $\times$ 2 cm.  
The timing and d$E/$d$x$ resolutions was 280 ps and 15 \%, respectively.
A detailed description of '93 instrument has been published 
elsewhere~\cite{BESS}.

\widetext

\narrowtext
These detector-components along with  electronics
are enclosed by a 2.5-mm-thick aluminum vessel that keeps the inside 
pressure at 1020 mb during the flights.
The total material-thickness along the particle path is 17.8 g/cm$^{2}$,
of which 8.6 in front of the central tracker, 0.4 between IDC and JET,
and 0.14 in the JET chamber gas, all in g/cm$^{2}$.

The trigger system is designed to detect negatively charged particles 
($\bar{p}$, antihelium) efficiently 
while sampling the proton and helium events in an unbiased manner.
This system is composed of the ``T0 trigger'' and the ``track trigger''. 
The T0 trigger initiates the data gathering based upon
a simple coincidence of the top and bottom scintillators with
thresholds set at 1/3 of the pulse height from a minimum
ionizing particle.
The track trigger consists of two sequential processes.
The first is a pattern-selection process to reject the null-track and 
multi-track events by using the hit-pattern of the IDC and ODC cells. 
If an event passes the pattern-selection, the second process performs a rough 
rigidity-selection by using the cell hit-pattern.
Irrespective of the track trigger condition, a fraction of T0 triggers
is recorded to provide an ``unbiased trigger sample'',
which is used to determine the efficiency of the track trigger selections. 
The data were collected in balloon-flights 
performed on July 26 '93, July 31 '94, and July 25 '95,
all launched from Lynn Lake, Manitoba, Canada.
During each level flight, the floating altitude was 36.5 km 
(residual atmospheric depth of 5 g/cm$^2$).
The total live time of the three flights was 84446 sec.

The following off-line selections 
are applied for both negative and positive
curvature events.
(i) One and only one track should be found in the JET chamber.
(ii) The track should be fully contained in 
the central six-column of the JET chamber 
(see fig~\ref{fig:bess}),
where the resolution of the rigidity-measurement is optimum.
This fiducial cut determines the geometrical acceptance 
for the present analysis 
of 0.35 m$^{2}$sr for '93 and '94, and 0.28 m$^{2}$sr for '95.
The following cuts are then applied to ensure the quality of the 
track and the timing measurement.
(1) The fitted $r\phi$-track should contain at least 14 hits in the
JET chamber and at least one hit in each of the upper two and 
lower two IDC layers.
(2) The reduced chi-square of the $r\phi$-track-fitting has to be less than 3.
(3) The extrapolated track should cross the fiducial region 
of the TOF scintillators
($|z| < 49.5$ cm for '93 and '94, $|z| < 47.0$ cm for '95). 
(5) The $z$-position ($z_{\text{TOF}}$) determined by the left-right 
time difference measured by the PMTs 
should match the $z$-impact point of the extrapolated track 
at the TOF counter within 10 cm ($|z_{\text{TOF}}-z|<10$ cm).

\begin{figure}[t]
\centerline{\mbox{\epsfxsize=9cm 
                  \epsffile{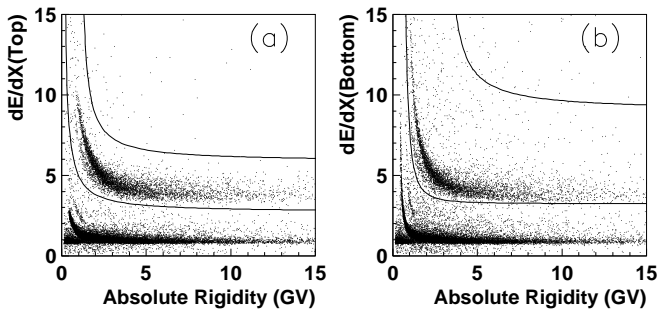}}}
\caption{Scatter plot of d$E/$d$x$ vs. absolute rigidity for 
(a) the top and (b) bottom layers of the TOF hodoscope ('95).
The solid lines show the d$E/$d$x$ band cut positions.}
\label{fig:dedx}
\end{figure}

\begin{figure}[tbp]
\centerline{\mbox{\epsfxsize=7cm 
                  \epsffile{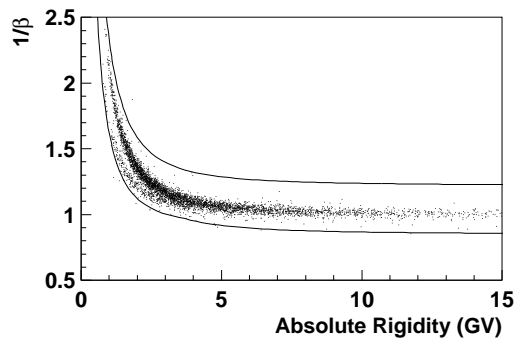}}}
\caption{Scatter plot of 1/$\beta$ vs. absolute rigidity.
The solid lines show the cut positions. ('95)}
\label{fig:beta}
\end{figure}

The particle velocity, i.e., $\beta(\equiv v/c)$ is 
determined from the time of flight and the path-length.  
Upward-moving particles are completely separated from 
downward-moving ones by the sign of $\beta$ due to the excellent 
time resolution.
We reject all upward-moving (albedo) particles at this stage, 
and limit further analysis to the downward-moving particles.
Fig.\ \ref{fig:dedx} (a) and (b) show plots of d$E/$d$x$ versus $|R|$ 
for the top and bottom scintillators, respectively,
for a sample of '95 data after applying all of the above mentioned cuts.
In order to select helium and antihelium,
we require that events must reside in the ``d$E/$d$x$ bands'' shown in 
Fig.\ \ref{fig:dedx} (a) and (b).
These d$E/$d$x$ cuts reject most of singly-charged particle events
(\prt, \pb, $e^{\pm}$), 
while preserving a high efficiency (95 \%) for helium.
Fig.\ \ref{fig:beta} shows plot of $1/\beta$ versus $|R|$
for a sample of '95 data after these d$E/$d$x$ cuts, where clear lines of 
$^4$He and $^3$He are visible. 
We further require that antihelium as well as helium
have $\beta$ and rigidity values inside the band shown 
in Fig.\ \ref{fig:beta}.  
\vspace{ 8 cm } 
The efficiency of the $\beta$-band cut is 99 \% for helium.
The overall efficiency of the off-line helium selection (from cut (1) to 
the $\beta$-band cut),
is 63 \%, 70 \%, and 78 \%, for '93, '94, and '95, respectively, and
71 \% in average (weighted by number of events).

\begin{figure}[tbb]
\centerline{\mbox{\epsfxsize=8cm 
                  \epsffile{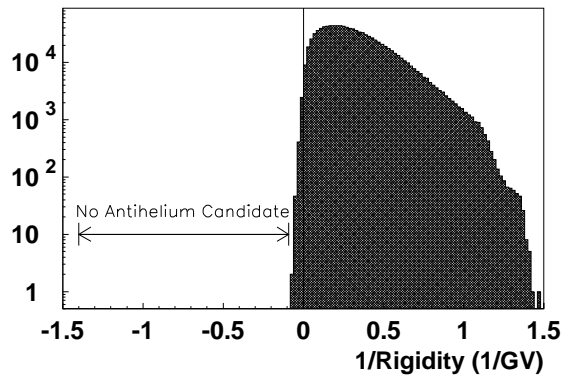}}}
\caption{ 1/rigidity distribution of the events which survive all selections.}
\label{fig:rgdty} 
\end{figure}

All selections up to this level use $|R|$ instead of $R$, i.e. do not 
discriminate the positive and negative rigidity (helium and antihelium).
Fig.\ \ref{fig:rgdty} shows the $1/R$ distribution of the events
which survived all the cuts mentioned above. 
Those events in the negative region are apparently
the spillover from the positive region due to the 
finite $1/R$ resolution. 
There exist no events to the left of -0.0625 GV$^{-1}$, which corresponds
to the rigidity of -16 GV. 
To define the lowest end of the rigidity region, we note that the number of
helium in Fig.\ \ref{fig:rgdty} drops sharply at 1.4 GV$^{-1}$, 
which corresponds
to 1 GV at the top of the atmosphere (TOA) after correcting for the
energy losses in the instrument and in the air.
Therefore, we conclude that we observed no antihelium in the rigidity region
from 1 to 16 GV at TOA.

The corresponding limit on the \Heb/He flux ratio at TOA can be obtained
by dividing the upper limit on the number of antihelium 
($\overline{\rm N}_{\rm lim}$) by the total number of helium (${\rm N}$),
which can be expressed as the integral; 
${ \rm N } = \sum_{i} [ {\rm n}_{\rm i} / ( \epsilon_{\rm pat}
                                            \epsilon_{\rm sel}
                                            \eta )_{\rm i} ] / {\rm f}$.
The summation is over the 60 bins which divide the rigidity region
(1 to 16 GV), ${\rm n}_{\rm i}$ is
is the number of helium detected in the unbiased trigger sample
falling in the i-th rigidity bin,
$\epsilon_{\rm pat}$
and $\epsilon_{\rm sel}$ are the efficiencies, for the helium, 
of the track-pattern- and the off-line selections, respectively,
and $\eta$ is the probability 
of the helium surviving through the air and the instrument, 
all at the i-th rigidity bin. 
The pre-set sampling factor ${\rm f}$ of the unbiased trigger was
1/40 for '93, 1/15 for the first half and 1/30 for the last half 
of '94, and 1/20 for '95.
The efficiencies $\epsilon_{\rm pat}$
as well as $\epsilon_{\rm sel}$ 
can be directly determined by using the unbiased trigger sample.
To calculate the surviving probability $\eta$, we utilize the
``inelastic''~\cite{inel}
cross sections of helium incident on various target nuclei, i.e. 
$\sigma$(He, $A_{\rm t}$). 
Since the data on $\sigma$(He, $A_{\rm t}$) at GeV energies is 
relatively sparse, we obtain $\sigma$(He, $A_{\rm t}$) by starting with
$\sigma$(\prt, $A_{\rm t}$), 
given as a function of incident energy~\cite{let83:pro}, 
and by scaling it using ``the hard sphere with overlap'' model 
of nuclear interaction~\cite{model}, i.e. 
\begin{eqnarray}
\label{eq:cros}
\sigma(A_{\rm i},A_{\rm t}) \propto 
(A_{\rm i}^{1/3} + A_{\rm t}^{1/3} - 
0.71 \times (A_{\rm i}^{-1/3} + A_{\rm t}^{-1/3}) )^{2}
\end{eqnarray}
where $A_{\rm i}$ and $A_{\rm t}$ are the atomic weight of incident and target
nuclei.
This model is known to reproduce data on nuclear interactions for various
combinations of $A_{\rm i}$ and $A_{\rm t}$ including the proton~\cite{cros}. 
The resultant $\sigma$(He, $A_{\rm t}$) agrees (within an accuracy
of 10 \%) with measured helium cross sections on targets such as C or Al.
This also justifies our use of the same formula~(\ref{eq:cros}) to estimate
the unmeasured $\sigma$(\Heb, $A_{\rm t}$) cross sections
from the $\sigma$(\pb, $A_{\rm t}$) data~\cite{antp}.
 
When summed over the rigidity bins, 
the number of detected helium events divided
by the sampling factors ($\sum n_{i}/{\mbox{\rm f}}$) are
$2.3 \times 10^{5}$, $8.7 \times 10^{5}$, and  $8.6 \times 10^{5}$
for '93, '94, and '95, respectively, and
the total number of helium ${\rm N}$ is $5.1 \times 10^{6}$.

The upper limit on the number of antihelium ($\overline{\rm N}$) depends on 
its unknown rigidity spectrum,
since the efficiencies are functions of rigidity.
In order to obtain the most conservative limit, irrespective of the
antihelium spectrum, we define  
$ \overline{\rm N}_{\rm lim} = 3.0 / [ \overline{\epsilon}_{\rm pat}
                            \overline{\epsilon}_{\rm rig}
                            \overline{\epsilon}_{\rm sel} 
                            \overline{\eta} ]_{\rm min} $,
where 
3.0 is the 95 \% confidence level limit for zero-observed events, and 
$[ \overline{\epsilon}_{\rm pat}
         \overline{\epsilon}_{\rm rig}
         \overline{\epsilon}_{\rm sel} 
         \overline{\eta} ]_{\rm min}$ is the minimum value of the
         antihelium efficiency-product
in the rigidity region.
In calculating
$[ \overline{\epsilon}_{\rm pat}
         \overline{\epsilon}_{\rm rig}
         \overline{\epsilon}_{\rm sel} 
         \overline{\eta} ]_{\rm min}$, 
we take $\overline{\epsilon}_{\rm sel}=\epsilon_{\rm sel}$ and
$\overline{\epsilon}_{\rm pat}=\epsilon_{\rm pat}$ for each rigidity bin, 
since the pattern-selection in the track trigger as well as the off-line
selection does not depend on the sign of the track rigidity.
The efficiency of the rigidity-selection in the track trigger 
$\overline{\epsilon}_{\rm rig}$
can be directly obtained by using events with a negative rigidity track 
in the unbiased trigger sample, 
and is a monotonic function of the rigidity, having values of 0.95 at 1 GV 
and 0.85 at 16 GV.
The efficiency product $(\overline{\epsilon}_{\rm pat}
         \overline{\epsilon}_{\rm rig}
         \overline{\epsilon}_{\rm sel} 
         \overline{\eta})$ thus does not vary significantly 
	 (0.19 at 0.7 GV, 0.23 at 2 GV, and 0.20 at 16 GV).
We take the minimum value of 0.19 to calculate $\overline{\rm N}_{\rm lim}$.

\begin{figure}[tbp]
\centerline{\mbox{\epsfysize=9cm 
                  \epsffile{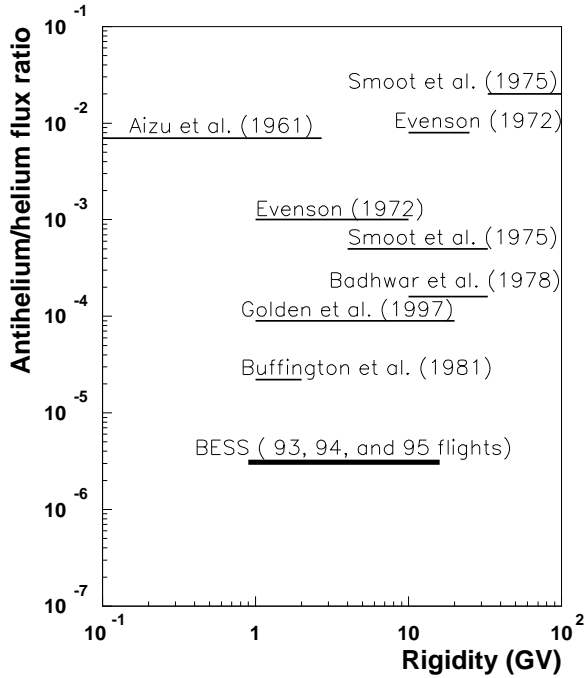}}}
\caption{
The resultant upper limit of \Heb/He flux ratio of the present paper 
together with previous limits.}
\label{fig:ratio}
\end{figure}

The resultant 95\% C.L. upper limit on the \Heb/He flux ratio at TOA 
is $3.1 \times 10^{-6}$ in the rigidity region from 1 to 16 GV.
The upper limit on the antihelium flux integrated over
the rigidity region is $6\times10^{-4}$ m$^{-2}$sr$^{-1}$s$^{-1}$.
It should be noted that these upper limits are very conservative and are 
valid for any hypothetical antihelium spectrum.
Our limit on the flux ratio obtained by this work 
is shown in Fig.\ \ref{fig:ratio} together with
the previous ones~\cite{EV72,PREV}.  Our flux ratio
is a factor 7 improvement over the limit of Buffington
et al. who looked for the annihilation signal of low rigidity
(1 $\sim$ 2 GV) antihelium in a spark chamber calorimeter,
and is 30 times more stringent than the limit of Golden et al. 
who covered a rigidity region similar to ours by using a previous
generation magnetic rigidity spectrometer.
The large acceptance of BESS, which is designed~\cite{SO87} specially 
for the antimatter and antiproton searches, made it possible to set
this limit.

Authors thank NASA/GSFC/WFF Balloon office and NSBF
for the balloon expeditions, and KEK and ISAS for various supports.
Sincere thanks are given to Y.~Ajima, Y.~Higashi and  
D.~Righter for their help 
with the flights and recovery, and to A. Moiseev for discussions.
This experiment was supported by NASA in the USA, and
by Grant-in-Aid for Scientific Research and  
for International Scientific Research
from the Ministry of Education, by the Kurata Research Grant, 
and by Sumitomo Research Grant in Japan. 
The analysis was performed on the
RS/6000 workstations supplied for the partnership program between ICEPP
and IBM Japan, Ltd.


\begin{references}
\bibitem{resol} All resolutions quoted in this paper are r.m.s. resolutions.
\end{references}
\end{document}